\documentstyle[12pt]{article}

\newcommand{\matel}[3]{\langle #1|#2|#3\rangle}
\newcommand{\ra}{\rightarrow}

\newcommand{\aver}[1]{\langle #1\rangle}

\newcommand{\al}{\alpha}
\newcommand{\as}{\alpha_s}

\newcommand{\ga}{\gamma}

\newcommand{\MeV}{\,\mbox{MeV}}

\newlength{\dinwidth}
\newlength{\dinmargin}
\begin{document}
{}~~\\
\vspace{1cm}
\begin{flushright}
\large{UND-HEP-95-BIG01\\
January 1995}
\end{flushright}
\begin{center}
\begin{Large}
\begin{bf}
%
% TITLE
%
A QCD TREATMENT OF THE WEAK DECAYS OF HEAVY FLAVOUR
HADRONS --
WITHOUT VOODOO AND UNDUE INCANTATIONS\footnote{Invited
Talk given at the
International Workshop on B Physics, Nagoya, Japan, Oct. 26 - 28,
1994}\\
\end{bf}
\end{Large}
\vspace{5mm}
\begin{large}
%
% Your NAME
%
I.I. Bigi\\
\end{large}
%
% Your AFFILIATION and ADDRESS
%
Physics Dept., University of Notre Dame du Lac\\
Notre Dame, IN 46556, U.S.A.\\
e-mail address: BIGI@UNDHEP.HEP.ND.EDU
\vspace{5mm}
\end{center}
%\begin{quotation}
\noindent
\begin{abstract}
There now exist several theoretical technologies for treating weak
decays of heavy flavour hadrons that are
genuinely based on QCD without having to invoke a deus ex machina.
I focus on one of those, which employs an expansion in inverse
powers
of the heavy quark mass. It has developed into a rather mature
framework
incorporating many subtle aspects of quantum field theory. I
describe its methodology for treating fully integrated decay rates as
well as differential distributions, in particular energy spectra; the
importance of a new type of sum rules, the SV sum rules, is
emphasized. First practical benefits from this theoretical technology
are listed, like predictions on lifetime ratios and extracting the KM
parameter $|V(cb)|$ from inclusive and from exclusive semileptonic
$B$ decays. An outlook is given onto future developments concerning
the determination of the properly defined mass of the heavy quark
and its kinetic energy and a reliable extraction of $|V(ub)/V(cb)|$. A
few comments on charm decays are added.
\end{abstract}
%
% SECTION I
%
\section{Introduction}
Early attempts to establish mastery over a complex problem often
contain
elements of Voodoo. Yet we have reached a stage where we can treat
the
weak decays of heavy flavour {\em hadrons} without such a time-
honoured
tool; our description can be based directly on QCD, with only one or
two
permissible incantations, as indicated later on. It is a three-fold
message I
want to give here: {\em significant progress} is being made which is
of considerable
{\em intellectual interest} and at the same time of increasing
{\em practical value}, most crucially for extracting KM parameters
from data. This in turn
will allow us to make Standard Model (SM) predictions more precise
both
parametrically and numerically and thus put searches for
New Physics (NP) onto a firmer basis.

Discovering CP asymmetries is seen as the ultimate prize in beauty
physics.
Those asymmetries being {\em linear} in the ratio of (coherent)
amplitudes
possess a high sensitivity to NP. Since the large samples of beauty
hadrons expected to be accumulated at the $e^+e^-$ $B$ factories and
ultimately
at the LHC would allow measurements with a statistical uncertainty
of a few
percent only, the question arises whether we can acquire sufficient
calculational control s.t. the basic quantities $|V(cb)|$ and $|V(ub)|$
can
be determined with at most a few percent and $|V(td)|$ with
10-15 \% uncertainty. Such objectives will not be attainable in the
very near
future or by theoretical advances alone. I anticipate it to be a rather
long-term project that involves an iterative feedback between
theoretical
analyses and a {\em broad} data base as its central elements. It
seems impossible
to predict on which particular route this project will proceed; yet
one can describe a few
promising avenues, identify several gateways and list
strategic elements to guide us in the future. To describe those
and illustrate them through possible itineraries is the goal of my talk:
in
Sect.2 I sketch the Heavy Quark Expansion for fully integrated
transition rates and describe some applications; in Sect.3 I extend
these
methods to make them applicable to differential distributions like
energy spectra, and in Sect.4 I will describe a new type of powerful
sum rules; in Sect.5 I discuss procedures for extracting
$|V(ub)/V(cb)|$ before presenting an outlook in Sect.6.

\section{$1/m_Q$ Expansions for Total Decay Widths}
\subsection{Methodology}
The weak decay of the heavy quark $Q$ inside the heavy flavour
hadron $H_Q$
proceeds within a cloud of light degrees of freedom (quarks,
antiquarks
and gluons) with which $Q$ and its decay products can interact
strongly. It
is the challenge for theorists to treat these initial and final state
hadronization effects. There are four Post-Voodoo theoretical
technologies available, namely QCD Sum Rules,
Lattice QCD, Heavy Quark Effective Theory (HQET) and Heavy Quark
Expansions. I will
mainly focus on the last technique, but also indicate where there is
an
overlap between different technologies, with an opportunity for
cooperation.

In analogy to the treatment of
$e^+e^-\rightarrow hadrons$ one describes the transition rate into an
inclusive final state $f$ through the imaginary part of a
forward scattering operator evaluated to second order in the weak
interactions \cite{BUV,BS,SV}:
$$\hat T(Q\rightarrow f\rightarrow Q)=
i\int d^4x\{ {\cal L}_W(x){\cal L}_W^{\dagger}(0)\} _T\eqno(1)$$
where $\{ .\} _T$ denotes the time ordered product and
${\cal L}_W$ the relevant effective weak Lagrangian expressed on
the
parton level. If the energy released in the decay is sufficiently large
one can express the {\em non-local} operator product in eq.(1) as an
infinite sum of {\em local} operators of increasing dimension with
coefficients containing higher and higher inverse powers of the
heavy quark mass $m_Q$.\footnote{It should be kept in mind,
though, that
it is primarily the energy release rather than $m_Q$ that controls the
expansion.} The width for $H_Q\rightarrow f$ is then obtained by
taking the
expectation value of $\hat T$ between the state $H_Q$; through
order
$1/m_Q^3$ one finds:
$$\Gamma (H_Q\ra f)=\frac{G_F^2m_Q^5}{192\pi ^3}|KM|^2
\left[ c_3(f)\frac{\matel{H_Q}{\bar QQ}{H_Q}}{2M_{H_Q}}+
\frac{c_5(f)}{m_Q^2}\frac{
\matel{H_Q}{\bar Qi\sigma _{\mu \nu}G_{\mu
\nu}Q}{H_Q}}{2M_{H_Q}}+ \right.
$$
$$\left. +\sum _i \frac{c_6^{(i)}(f)}{m_Q^3}\frac{\matel{H_Q}
{(\bar Q\Gamma _iq)(\bar q\Gamma _iQ)}{H_Q}}
{2M_{H_Q}} + {\cal O}(1/m_Q^4)\right] ,  \eqno(2)$$
where the dimensional coefficients $c_i(f)$ depend on the parton
level
characteristics of $f$ (such as the ratios of the final-state quark
masses
to $m_Q$), $KM$ denotes the appropriate combination of KM
parameters,
and $G_{\mu \nu}$ the gluonic field strength tensor. The last term
implies also the summation over the four-fermion operators with
different light flavours $q$. The factor $1/2M_{H_Q}$ reflects the
relativistic normalization of the state $|H_Q\rangle$. It is through
$\matel{H_Q}{O_i}{H_Q}$, the expectation values of the local
operators
$O_i$, that the dependence on the {\em decaying hadron} $H_Q$, and
on
non-perturbative forces in general, enters. Since these are matrix
elements
for on-shell hadrons $H_Q$, one sees that $\Gamma (H_Q\ra f)$ is
indeed
expanded into a power series in $\mu _{had}/m_Q$.

Four general remarks are in order at this point:

\noindent (A) At first one might think that the $1/m_Q$ scaling
sketched
above is vitiated by gluon radiation. Yet it persists to hold
for fully inclusive transitions \cite{MIRAGE}.

\noindent (B) The most important element of eq.(2) is -- the one that
is
{\em missing}! Namely there is no term of order $1/m_Q$ in the
total decay width whereas such a correction definitely exists for the
mass formulae -- $M_{H_Q}=m_Q(1+\bar \Lambda /m_Q+{\cal
O}(1/m_Q^2))$ -- and likewise for differential decay distributions, to
be discussed later.
It can be shown that integrated widths are free from $1/m_Q$
corrections
due to a delicate cancellation between initial and final state
hadronization
effects as imposed by local colour symmetry. This can be understood
in
another more compact (though less intuitive) way as well: with the
leading operator $\bar QQ$ carrying dimension three only
dimension four operators can generate
$1/m_Q$ corrections; yet there is no independent dimension four
operator \cite{CHAY,BUV}
{\em once the equation of motion is imposed} -- unless one abandons
local
colour symmetry thus making the operators $\bar Qi\gamma \cdot
\partial Q$
and $\bar Qi\gamma \cdot G Q$ independent of each other
($G_{\mu}$ denotes the gluon field)! The leading
non-perturbative corrections to fully integrated decay widths are
then
of order $1/m_Q^2$ and their size is controlled by two
dimension five operators, namely the chromomagnetic and the
kinetic
energy operators.

\noindent (C) Because non-perturbative corrections to
total widths are of order
$1/m_Q^2$ rather than $1/m_Q$, they amount to no more than 10
percent
for $B$ mesons: $(\mu _{had}/m_b)^2$ $\simeq ({\cal O}(1\,
GeV)/m_b)^2
\sim {\cal O}(\% )$ (details will be given below).
To predict the decay rates for
beauty hadrons as a function of $V(cb)$ or $V(ub)$ with a theoretical
uncertainty of less than a few percent -- which is our goal --we
therefore have to establish computational control
over the non-perturbative corrections merely on the $\sim 20\%$
level.  As a side remark: one concludes likewise that the decays of
$B_c$ mesons are mainly driven by the decays of charm
rather than beauty quarks inside the $B_c$
resulting in a lifetime considerably shorter than 1 psec.
The situation is obviously numerically ambivalent for the decays of
charmed hadrons; in the following I will concentrate on beauty
decays with just a few remarks on charm decays.

\noindent (D) For a $1/m_Q$ expansion it is of course important
to understand which kind of quark mass is to be employed there,
in particular since
for confined quarks there exists no a priori natural choice.
It had been claimed that the pole mass can and therefore
should conveniently be used. Yet such claims
turn out to be fallacious \cite{POLEMASS,BRAUN}: QCD, like QED, is
not
Borel summable; in the high order terms
of the perturbative series there arise instabilities
which are customarily referred to as
(infrared) renormalons representing poles in the Borel plane; they
lead to an {\em additive} mass renormalization generating an
irreducible uncertainty
of order $\bar \Lambda$ in the size of the pole mass:
$m_Q^{pole}=m_Q^{(0)}(1+c_1\as + c_2\as ^2 +...+ c_N\as ^N) +
{\cal O}(\bar \Lambda )=m_Q^{(N)}(1+{\cal O}(\bar \Lambda/
m_Q^{(N)}))$.   While this effect
can safely be ignored in a purely perturbative treatment, it negates
the
inclusion of non-perturbative corrections $\sim {\cal O}(1/m_Q^2)$,
since those are
then parametrically smaller than the uncertainty in the definition of
the
pole mass. This problem can be taken care of through
Wilson's prescriptions for the operator product expansion:
$$\Gamma (H_Q\ra f)=\sum _i c_i^{(f)}(\mu )
\matel{H_Q}{O_i}{H_Q}_{(\mu )}\eqno(3)$$
where a momentum scale $\mu$ has been introduced to allow a
consistent
separation of contributions from Long Distance and Short Distance
dynamics: $LD > \mu ^{-1} > SD$ with the latter contained in
the coefficients $c_i^{(f)}$ and the former lumped into the matrix
elements. The quantity $\mu$ obviously represents an auxiliary
variable which drops out from the observable, in this case the decay
width. In the limit
$\mu \ra 0$ infrared renormalons emerge in the coefficients; they
cancel
against ultraviolet renormalons in the matrix elements, yet that does
{\em not} mean that these infrared renormalons are irrelevant and
that
one can conveniently set $\mu =0$! For to incorporate both
perturbative
as well as non-perturbative corrections one has to steer a careful
course
between `Scylla' and `Charybdis': while one wants to pick
$\mu \ll m_Q$ so as to make a heavy quark expansion applicable,
one
also has to choose $\mu _{had}\ll \mu$ s.t. $\alpha _S(\mu ) \ll 1$;
for otherwise the {\em perturbative} corrections become
uncontrollable. Wilson's
OPE allows to incorporate both perturbative and non-perturbative
corrections, and {\em this underlies also a consistent formulation of
HQET};
the scale $\mu$ provides an infrared cut-off that automatically
freezes out infrared renormalons. For the asymptotic difference
between the hadron and the quark mass one then has to write
$\bar \Lambda (\mu ) \equiv
(M_{H_Q}-m_Q(\mu ))_{m_Q\ra \infty}$.
This nice feature does not come for free,
of course: for one has to use a `running' mass $m_Q(\mu )$
evaluated at an
intermediate scale $\mu$ which presents some technical
complication. If there
exists a framework other than Wilson's OPE to do the trick, I would
be
most eager to hear about it.

Next I want to address the question of how to
determine the various expectation values. Using heavy quark
expansions one can relate some of these matrix elements -- and in
particular
those appearing in the leading terms in eq.(2) -- to other observables
to
extract their size. We will also see that it is here where a very
fruitful
cooperation with analyses based on QCD Sum Rules and on Lattice
QCD is
emerging. Using the equation of motion one finds
$$\matel{B}{\bar bb}{B}/2M_B=1-\frac{\langle (\vec p_b)^2\rangle
_B}
{2m_b^2}+\frac{3}{8}\frac{M^2_{B^*}-M^2_B}{m_b^2}+{\cal
O}(1/m_b^3)
\eqno(4)$$
where $\langle (\vec p_b)^2\rangle _B\equiv \matel{B}{\bar b(i\vec
D)^2b}{B}
/2M_B$ describes the motion of the $b$ quark under the influence of
the
gluon background field inside the $B$ meson ($\vec D$ denotes the
covariant derivative); here I have already invoked
heavy quark symmetry to obtain:
$$\matel{B}{\bar bi\sigma \cdot Gb}{B}/2M_B\simeq
\frac{3}{2} (M^2_{B^*}-M^2_B)\simeq 0.74\, (GeV)^2\eqno(5)$$
For baryons $\Lambda _b$ one has
$\matel{\Lambda _b}{\bar b i\sigma \cdot Gb}{\Lambda _b}\simeq
0$.
The difference
$\langle (\vec p_b)^2\rangle _{\Lambda _b}-
\langle (\vec p_b)^2\rangle _B$ can be deduced from the hadron
masses \cite{BUVPREPRINT}:
$$\langle (\vec p_b)^2\rangle _{\Lambda _b}-
\langle (\vec p_b)^2\rangle _B\simeq \frac{2m_bm_c}{m_b-m_c}
\{ \langle M_B\rangle -M_{\Lambda _b}-
\langle M_D\rangle +M_{\Lambda _c}\}
\simeq -(0.07 \pm 0.20)\, (GeV)^2 \eqno(6)$$
using present data; $\langle M_{B,D}\rangle$ denote spin averaged
meson masses. The size of
$\langle (\vec p_b)^2\rangle _B$ is not precisely known yet, beyond
the inequality \cite{VOLOSHIN,OPTICAL}
$$\langle (\vec p_b)^2\rangle _B\geq \langle \mu ^2_G\rangle
_B\equiv
\frac{1}{2}\matel{B}{\bar bi\sigma \cdot Gb}{B}/2M_B\simeq 0.37\,
(GeV)^2\eqno(7)$$
which is almost saturated by the QCD sum rules estimate
\cite{QCDSR} of around
$0.5\, (GeV)^2$. Even present day lattice QCD should be able
to extract $\langle (\vec p_b)^2\rangle _B$ from determining
$\langle (\vec p_c)^2\rangle _D$ and varying the charm mass $m_c$
\cite{GUIDO}; likewise for
$\matel{B(p)}{(\bar b_L\gamma _{\mu}q_L)(\bar q_L\gamma
_{\mu}b_L)}{B(p)}$.
Thus we find for the expansion parameters:
$$\frac {\langle (\vec p_b)^2\rangle _B}{m_b^2} \simeq 0.016
\simeq \frac{\langle \mu ^2_G\rangle _B}{m_b^2}\; \; ,
\; \frac {\langle (\vec p_b)^2\rangle _D}{m_c^2} \simeq 0.21
\simeq \frac{\langle \mu ^2_G\rangle _D}{m_c^2}\eqno(8)$$
for beauty and charm, respectively.
We also have
$$m_b-m_c\simeq \langle M\rangle _B - \langle M\rangle _D +
\langle (\vec p)^2\rangle \cdot
\left( \frac{1}{2m_c}- \frac{1}{2m_b}\right)
\simeq 3.46 \pm 0.04\, GeV\, .\eqno(9)$$
One should note that this mass {\em difference} is free of renormalon
contributions.

\subsection{Applications}
With the theoretical elements thus assembled I can discuss four
applications.
\subsubsection{Lifetime Ratios}
The predictions for the lifetime ratios of beauty
hadrons are given in
Table \ref{TABLE1}:
\begin{table}
\begin{tabular} {|l|l|l|}
\hline
%& &\\
Observable &QCD ($1/m_b$ expansion) &Data \\
%& & \\
\hline
\hline
%&  & \\
$\tau (B^-)/\tau (B_d)$ & $1+
0.05(f_B/200\, \MeV )^2
[1\pm {\cal O}(10\%)]>1$ &$0.98 \pm 0.09$ \\
&(mainly due to {\em destructive} interference) & \\
\hline
%& & \\
$\bar \tau (B_s)/\tau (B_d)$ &$1\pm {\cal O}(0.01)$
&  $ 0.89\pm 0.20$ \\
%& & \\
\hline
%& & \\
$\tau (\Lambda _b)/\tau (B_d)$&$\sim 0.9$ & $0.71\pm 0.14$\\
%& & \\
\hline
\end{tabular}
\centering
\caption{QCD Predictions for Beauty Lifetime Ratios}
\label{TABLE1}
\end{table}
%\vspace{0.5cm}
There is an apparent -- though not yet conclusive -- discrepancy
between the
theoretical expectation and the present data on $\tau (\Lambda
_b)/\tau (B_d)$.
Therefore I want to comment briefly on the theoretical foundations
for that
prediction. Applying $1/m_Q$ expansions properly defined in
Euclidean space
to the decays of real hadrons $H_Q$ proceeding in
Minkowskian space requires certain assumptions concerning the
analyticity structure of a transition amplitude; those transcend what
today can
rigorously be proven within QCD. Yet I consider that to constitute
merely an
allowed incantation since under the concept of `quark-hadron
duality' it
represents an integral part of practically all applications of QCD.
While
it is conceivable that duality holds for semileptonic, but not for
non-leptonic beauty decays, we have been unable to discern any
reason for
such a selective qualitative failure of duality. On the other
hand, it is quite possible that the $1/m_Q$ expansion, which is
actually controlled by the energy release per quark, converges more
slowly
for $b\ra c\bar ud,\, c\bar cs$ than for $b\ra l\nu c/u$ or even
$b\ra u\bar ud$. Personally I am concerned about the short
$\Lambda _b$ lifetime; yet I do not consider it a theoretical
desaster -- unless $\tau (\Lambda _b)/\tau (B_d)\simeq 0.5$ were
to hold -- and I am inclined to rely on the last resort available to
`catholic' reasoning, namely to light a candle in church and pray for
ultimate redemption.

The expectations \cite{MIRAGE,MARBELLA,DS}
for the lifetime ratios of charm hadrons are juxtaposed to  the data in
Table \ref{TABLE2}:
\begin{table}
\begin{tabular} {|l|l|l|}
\hline
%& &\\
Observable &QCD ($1/m_c$ expansion) &Data \\
%& & \\
\hline
\hline
%&  & \\
$\tau (D^+)/\tau (D^0)$ & $\sim 2$ &$2.50 \pm 0.05$ \\
\hline
%& & \\
$\tau (D_s)/\tau (D^0)$ &$1\pm$ few \%
&  $ 1.13\pm 0.05$ \\
%& & \\
\hline
%& & \\
$\tau (\Lambda _c)/\tau (D^0)$&$\sim 0.5$ & $0.51\pm 0.05$\\
%& & \\
\hline
$\tau (\Xi ^+ _c)/\tau (\Lambda _c)$&$\sim 1.3$ & $1.68\pm 0.5$\\
%& & \\
\hline
$\tau (\Xi ^+ _c)/\tau (\Xi ^0 _c)$&$\sim 2.8$ & $2.46\pm 0.75$\\
\hline
\end{tabular}
\centering
\caption{QCD Predictions for Charm Lifetime Ratios}
\label{TABLE2}
\end{table}
%\vspace{0.5cm}
The agreement with the data is surprisingly good for a $1/m_c$
expansion; it also should be noted that quark model estimates were
used for some of the
{\em baryonic} expectation values. I will, however, present an
observation below which might suggest that this agreement is
somewhat coincidental.

\subsubsection{Semileptonic Branching Ratio of B Mesons}
The present world average yields \cite{CLEO}
$$BR_{SL}(B)\equiv BR(B\ra l\nu X)= 0.1043 \pm 0.0024
\eqno(10)$$
making it unlikely that $BR_{SL}(B)$ actually exceeds $0.11$ in any
significant way. A free parton model treatment leads to
$BR_{SL}(b)|_{PM}\simeq 0.15$; inclusion of perturbative QCD lowers
it:
$BR_{SL}(b)|_{pert.QCD}\simeq 0.125 - 0.135$ \cite{PETRARCA}. The
data
differ from this expectation by $\sim 15 - 20\%$. Originally there
was no
clear need to view this difference as alarming; for a priori one would
think that the non-perturbative corrections transforming
$BR_{SL}(b)$ into
$BR_{SL}(B)$ could naturally close the gap since they would be of
order $\mu _{had}/m_b\sim 10-20\%$ for $\mu _{had}\sim 0.5-1\,
GeV$. Yet we know now that the leading non-perturbative
contributions arise only at the level of $(\mu _{had}/m_b)^2\sim 1-
4\%$. A
more detailed analysis shows \cite{BUV}
that $BR_{SL}(B)$ is indeed lowered relative
to $BR_{SL}(b)$, but only by $\sim\, 2\%$. There exists a loophole,
though, in
that analysis: the energy release in the channel $b\ra c\bar cs$ is not
large
and corrections that are formally of order $1/m_b^3$ and higher
might
actually be numerically quite significant there. There is some
theoretical evidence
that they would indeed enhance $\Gamma (B\ra [c\bar cs])$. If
$\Gamma (B\ra [c\bar cs\bar q])\simeq 2\cdot \Gamma (b\ra c\bar
cs)$
were to hold, the non-leptonic $B$ width would be enhanced
sufficiently to bring the prediction on $BR_{SL}(B)$ in line with the
data. There is one problem with such a resolution: it would raise the
charm content $N_c$ of the decay products quite significantly:
$$N_c=1.3\;  \; \; \; \; if \; \; \; \;
\Gamma (B\ra c\bar cs\bar q)\simeq 2\cdot
\Gamma (b\ra c\bar cs)\eqno(11a)$$
$$N_c=1.15\;  \; \; \; \; if \; \; \; \;
\Gamma (B\ra c\bar cs\bar q)\simeq
\Gamma (b\ra c\bar cs)\eqno(11b)$$
Data are below, yet still compatible with $N_c=1.15$, but not with a
high value of 1.3.

To summarize the present situation \cite{BAFFLING}:
there is little doubt left that
experimentally $BR_{SL}\simeq 0.10-0.11$ indeed holds
while the question
of the charm content of the final state still remains somewhat
unsettled; on the theoretical side it would be quite premature to
invoke the lower than expected value for $BR_{SL}(B)$ as
evidence for New Physics in non-leptonic $B$ decays; on the other
hand the prediction is still above the measured value although there
are
indications that some higher order perturbative contributions are
larger
than originally thought and they reduce $BR_{SL}(b)$ somewhat
\cite{NIERSTE}. For proper
perspective one should also keep in mind that the absolute size of
the semileptonic branching ratio can be predicted with less precision
than
the {\em ratio} of semileptonic branching ratios and of lifetimes. For
it --
in contrast to the latter -- receives both perturbative and non-
perturbative
contributions and a precise numerical separation of the two is not
an easy task.

\subsubsection{$B\ra \gamma +X_{s,d}$ Transitions}
Through order $1/m_b^2$ one finds \cite{MANIFESTO}
$$\Gamma (B\ra \gamma X_q)=\Gamma (b\ra \gamma q)
\left( \frac{\matel{B}{\bar bb}{B}}{2M_B} + f\left(
\frac{m_q^2}{m_b^2}\right)
\frac{\aver{\mu ^2_G}_B}{m_b^2}
+{\cal O}(1/m_b^3)\right) \eqno(12)$$
for $q=s,d$ with $f(m_q^2/m_b^2)$ representing a phase space factor
and
therefore:
$$\frac{\Gamma (B\ra \gamma X_d)}{\Gamma (B\ra \gamma
X_s)}\simeq
\frac{|V(td)|^2}{|V(ts)|^2} + {\cal O}(1/m_b^3)\eqno(13)$$
This result is cute, yet at the same time quite useless. For it is due to
the
fact that the
difference between $m_s$ and $m_d$ can be ignored on the scale of
$m_b$; yet
by the same token there is no effective kinematical distinction
between
$X_s$ and $X_d$ final states; on the other hand fully reconstructing
them
is not a realistic proposition.

At (formal) order $1/m_b^3$ also a serious {\em theoretical}
problem emerges:
there is a {\em non-local} dimension eight operator
\footnote{The underlying process
can be described by a diagram where a $W$ is exchanged between
the $b$ quark
and the $\bar q$ antiquark after a photon is emitted from the $\bar
q$ line.
There are contributions that cancel against these terms
\cite{MIRAGE},
yet they represent
electromagnetic corrections to non-leptonic $B$ decays and have to
be counted there.}
generating contributions
dominated by long distance dynamics, in particular to $B\ra \gamma
X_d$; those
depend on $|V(ub)|$ rather than on $|V(td)|$ (or $|V(ts)|$) and at
present
cannot be evaluated in a reliable fashion.
A fortiori one has to be concerned that long distance
dynamics will affect in particular also the exclusive channels
$B\ra \gamma \rho , \, \gamma \omega$ \cite{SONI}.
Before measurements of
$BR(B\ra \gamma \rho , \, \gamma \omega)/BR(B\ra \gamma K^*)$
can be used to determine $|V(td)/V(ts)|$, one has to remove
theoretically these
non-Penguin contributions. The only reliable way I know of for
achieving that
is to measure $BR(B\ra \gamma D^*)$ and/or $BR(D\ra \gamma K^*)$
which
do not receive any Penguin contributions; these amplitudes can then
be used to gauge the size of the corresponding quantities in
$B\ra \gamma \rho /\omega$ and subtract them there. The left-over
part represents the Penguin contribution.

\subsubsection{Extracting $|V(cb)|$ from $\Gamma _{SL}(B)$
(and $m_c$ from $\Gamma _{SL}(D))$}

The semileptonic width of $B$ mesons can be expressed as follows:
$$\Gamma _{SL}(B)=\frac{G_F^2m_b^5}{192\pi ^3}|V(cb)|^2 \cdot $$
$$\left(
[z_0(x)-\frac{2\al _S}{3\pi}(\pi ^2-\frac{25}{4})z_1(x)]
[1-\frac{\aver{(\vec p_b)^2}_B -
\aver{\mu _G^2}_B}{2m_b^2}]-
z_2(x)\frac{\aver{\mu _G^2}_B}{m_b^2}+
{\cal O}(\al _S^2,\frac{\al _S}{m_b^2},
\frac{1}{m_b^3})\right) \eqno(14)$$
where the $z_0,z_1$ and $z_2$ represent known phasespace
functions of $x=m_c^2/m_b^2$. To  appreciate
this formula one should note the following: the non-perturbative
corrections are small -- the leading ones arise only at order
$1/m_b^2$ -- and rather well known numerically:
for $\aver{\mu _G^2}_B$
is given by the meson hyperfine splitting, see eq.(5), and there
exist decent bounds on $\aver{(\vec p_b)^2}_B$. The main
numerical uncertainty
actually derives from the proper choice of a value for
the quark mass $m_b$. Since $m_b$ is raised to the fifth power it
would
appear that this problem introduces a large theoretical uncertainty
into any attempt to extract $|V(cb)|$ from the measured semileptonic
width. Yet heavy quark symmetry comes to the rescue here. It turns
out
that $\Gamma _{SL}(B)$ depends mainly on the difference $m_b-
m_c$ rather
than on $m_b$ and $m_c$ separately; $m_b-m_c$ is tightly
constrained by
the measured meson masses, see eq.(9). An independant cross check
is
provided by the observation that also the {\em shape} of the lepton
spectrum
is mainly controlled by $m_b-m_c$; its value can then be extracted
from the
data and is in full agreement with the value from eq.(8)
\cite{VOLOSHIN2}
\footnote{These two methods for extracting $m_b - m_c$ are quite
independant of each other. The fact that they yield a practically
identical
value shows that higher order corrections that have been ignored
here do not make anomalously large contributions.}. Using that
information we then find \cite{SUV}
$$|V(cb)|_{incl}\simeq (0.0410\pm 0.002)\cdot \sqrt{\frac{1.5\,
psec}{\tau _B}}
\cdot \sqrt{\frac{BR_{SL}(B)}{0.1043}}\eqno(15)$$
where the stated error is theoretical and reflects the remaining
uncertainty in the size of $m_b$ and $m_c$. Contrary to some recent
claims in the literature, the perturbative corrections are under
control when evaluated at the appropriate scale \cite{SU}.

Equating the observed width $\Gamma _{SL}(D)$ with its theoretical
expression (and assuming $|V(cs)|\simeq 1$) leads to the
requirement
$"m_c"\simeq 1.6$ GeV \cite{DIEKMANN,DS}. However this is a high
value relative to what is derived from charmonium spectroscopy,
namely $m_c\leq 1.4$ GeV. A
difference of $0.2$ GeV in $m_c$ might appear quite innocuous --
till one
realizes that the corresponding semileptonic width depending on
$m_c^5$ differs by a factor of two or more! Quite generally, one
might suspect higher-order perturbative and non-perturbative
corrections to be
large. The analysis of ref.(\cite{DIEKMANN})
finds however that they  show a strong tendency to further {\em
decrease}  $\Gamma _{SL}(D)$ and their inclusion thus
does not help at all
to reproduce the measured value of $\Gamma _{SL}(D)$
with $m_c\simeq 1.4$ GeV. At present two possible
conclusions can be drawn from this observation :
$1/m_c$ expansions do not provide a reliable guide for charm
decays since
quark-hadron duality is vitiated  due to `not-so-distant cuts' in
charm
decays \cite{DIEKMANN}; or they can be trusted -- and even then
only
with quite a grain of salt --
only for lifetime {\em ratios} or {\em ratios} of semileptonic
branching ratios.

\section{Energy Spectra}

At first it would seem that it is beyond the reach of $1/m_Q$
expansions to
describe energy spectra in beauty decays. The argument goes as
follows.
Consider for simplicity the photon spectrum in $B\ra \gamma X$. To
leading
order in $1/m_b$ this decay is described by the quark level
transition
$b\ra \gamma s$ where the $\gamma$ spectrum consists of a single
line located at $E_{\gamma}\simeq m_b/2$; gluon bremsstrahlung
off the $s$ quark generates a perturbative tail for
$E_{\gamma}<m_b/2$.
Actually for $E_{\gamma}$ close to $m_b/2$ this gluon radiation
does
not represent a perturbative phenomenon. In any case one
encounters
the following problem: no events can be generated  beyond the
quark level
kinematical boundary, i.e. with
$E_{\gamma}>m_b/2$. On the other hand the true kinematical
boundary is set by the higher hadron mass $M_B/2$. It would then
appear that
in such a treatment no events could be generated with a
photon energy $E_{\gamma}$
in the {\em window} $[m_b/2,M_B/2]$ --
in clear conflict with observation. It is
intuitively clear how this conflict gets resolved physically: the $b$
quark is not at rest inside the $B$ meson; its `Fermi' motion spreads
the
photon line out over a region $\sim \bar \Lambda \equiv M_B-m_b$
thus
populating also the window $[m_b/2,M_B/2]$. This intuitive parton
level
picture has first been introduced by the authors of ref.
\cite{ALI-PIET};
in ref.\cite{ACM} it has been refined by carefully imposing
energy-momentum conservation; I will refer to it as the $AC^2M^2$
model.
The important new element
\cite{MANIFESTO,PRL} (see also
ref.\cite{MANOHAR}) is that this intuitive physical picture can
also be realized within QCD in a rigorous fashion through a $1/m_Q$
expansion, albeit with a certain subtle, yet relevant qualification:
with $E_{\gamma}$ in the window region the mass of the produced
hadronic system is typically of order $\bar \Lambda m_b$, i.e.
(moderately) large; yet for
$m_b/2-\bar \Lambda ^2/2m_b\leq E_{\gamma} \leq m_b/2$, i.e.
that highest slice of the
endpoint region, this mass is of order $\bar \Lambda ^2$ only, and
one
cannot trust the result of the OPE, evaluated to a finite order. It is
also
clear that for $m_s=0$ the $1/m_b$ expansion is even singular
around the endpoint $m_b/2$; for only in that way can
non-vanishing contributions beyond the quark level kinematical
boundary emerge
in a heavy quark expansion. The expansion parameter is actually
$\mu _{had}/[(1-y)m_b]$ with $y=2E_{\gamma}/m_b$. For $y$ not
too
close to unity one can compute the spectrum directly. Through order
$1/m_b^2$ the non-perturbative corrections to the spectrum are
expressed in
terms of $\aver{(\vec p_b)^2}_B$ and $\aver {\mu _G^2}_B$.
Applying it to semileptonic $B$ decays one obtains a result that is
very
similar to a {\em fit} of the $AC^2M^2$ model to the data. This is
quite
remarkable since in principle there are no free parameters, although
in
practise there is some `wiggle space' in the numerical size of $m_b$
and
$\aver{(\vec p_b)^2}_B$. Furthermore the region $y\simeq 1$
represents a
`black box' rather than `terra incognita', since one knows the
spectrum
integrated over the endpoint region: with the
total semileptonic width obtained from a non-singular expansion in
$1/m_b$
and the lepton spectrum calculable for $y$ not too close to unity one
can
conclude that the spectrum integrated down from unity to such
values of
$y$ can be determined as well.

This can be expressed in a rather transparent way for
$B\ra \gamma X$. The finite spread of the photon spectrum is given
by a
series of $\delta ^{(n)}(1-y),\, n=0,1,2,...$, i.e. $\delta$ functions
and their derivatives as a singular expansion around the
quark level endpoint $E_{\gamma}=m_b/2$. The coefficient of the
$\delta ^{(n)}(1-y)$ term then represents the $(n+1)$th moment
of the spectrum; these moments can then be calculated as
expectation values
of local operators. I will return to this point when discussing the new
sum rules.

It has been shown that in QCD proper one can indeed define and
calculate a
function $F$ that describes the motion of the $b$ quark inside
beauty hadrons \cite{MOTION,NEUBERT}. Yet this function possesses
some subtle features:
(i) While the motion it describes is non-relativistic, it {\em cannot}
be expressed through a non-relativistic hadronic wavefunction; for
the
third and higher moments of $F$ depend on the expectation values of
{\em time} components of various operators. (ii) The nature of $F$
and
the way it is computed depend quite sensitively on the final state
quark mass $m_q$ (in $Q\ra q$). For light quark masses
-- $\bar \Lambda ^2 \sim m_q^2 \ll m_b^2$ -- one obtains $F(x)$
from a
light cone correlator with $x=2(E_{\gamma}-m_b/2)/\bar \Lambda$.
For
heavy quarks with $m_q^2 < m_b^2$ on the other hand one expresses
$F(x)$ as a temporal correlator with $x=(E_{\gamma}-E_0)/\bar
\Lambda$,
$E_0=m_b(1-m_q^2/m_b^2)/2$. That also means that the function
$F(x)$
extracted from the heavy quark case {\em cannot} be used literally
in describing
the light quark case. I will return to this point below. (iii) The
situation is further complicated by the fact that a third relevant
scenario
exists, namely for $m_q^2\sim \bar \Lambda m_b$; this is the case
for
charm quarks! In some ways charm quarks in the final state
can then superficially be described like light rather than heavy
quarks;
however, as I will discuss below, there exists strong circumstantial
evidence that charm quarks behave more like heavy quarks in
beauty
decays.

It has been shown \cite{ACMREFINED} that a {\em judiciously
redefined} $AC^2M^2$ model
implements QCD for $b\ra u$ (as in $b\ra l\nu u$) and $b\ra s$
(as in $b\ra s\gamma$) transitions in a very satisfactory way. It
does not do quite as well
for $b\ra l\nu c$  decays as discussed later.

{\em In summary:}

\noindent $\bullet$ The $1/m_Q$ expansion allows to compute
energy
spectra in inclusive semileptonic and radiative beauty decays in
terms of
expectation values of local operators. It provides a very satisfactory
description of the presently available data.

\noindent $\bullet$ Close to the kinematical endpoint one has to
evaluate an increasing number of such expectation values; thus there
exist at present practical limitations for computing the endpoint
spectrum
directly.

\noindent $\bullet$ Even a perfect fit to the lepton spectrum in
$B\ra l\nu X_c$ does not allow us {\em per se} to compute the
lepton spectrum for $B\ra l\nu X_u$ as a function of
$|V(ub)/V(cb)|$ {\em alone}.

\section{The SV Sum Rules}

Semileptonic or radiative decays of heavy flavour hadrons $H_Q$ can
be
viewed as the inelastic scattering of virtual $W$ bosons or photons
off an
$H_Q$ target. This general analogy had been recognized already in
the early days of heavy quark symmetry \cite{CHAY,BJ};  invoking
the concept of quark-hadron
duality in various forms several sum rules had been written down
equating moments of observable energy spectra in inclusive decays
with the corresponding quantities evaluated on the parton level
\cite{BJ,BJETAL,VOLOSHINSR,LIPKIN}. Very recently it
has been shown \cite{OPTICAL} that these sum rules and a host of
new ones can consistently be derived from QCD proper in a certain
limit to be explained below, and that this can be achieved by
pursuing the correspondence with deep-inelastic lepton-nucleon
scattering: the differential decay rate is first written down in terms
of (five) Lorentz-invariant functions; a $1/m_Q$ expansion allows to
express those functions through the $H_Q$ expectation values of local
operators of higher and higher dimension \cite{LEW}.  Forming
moments of these universal functions
by integrating judiciously over the energy projects out certain
expectation values. In deep-elastic lepton-nucleon scattering this
procedure allowed to compute the evolution of such moments as a
function of the momentum
transfer. In heavy flavour decays one can harness heavy flavour
symmetry
to evaluate some of these matrix elements; this holds in particular if
one can
use the velocity of the hadronic system present in the decay as a
second expansion parameter, i.e. in the Small Velocity (SV) limit.

We thus find the following: the sum rules mentioned above which
seemed to
be unrelated to each other actually represent just different moments
of the same observable spectral distributions! Furthermore this
approach allows to
derive new sum rules as well as corrections to the previously found
ones in
a systematic and comprehensive fashion. There are numerous
benefits to
be derived from these sum rules; I will address here the following
ones:
(i) They provide valuable insights into the inner workings of quark-
hadron
duality. (ii) They enable us to deduce the numerical value for the
mass of the heavy quark and its kinetic energy from the data. (iii)
They allow us to derive the inequality of eq.(7) in a field-theoretic
manner. (iv) They provide us with a highly relevant bound on
$|F_{B\ra D^*}(0)|$, the formfactor for
$B\ra l\nu D^*$ at zero recoil.

{\em ad (i):)} For $B\ra \gamma X_h$, where $X_h$ denotes a heavy
state making the SV limit applicable here, one finds for the deviation
of the
photon energy from the zeroth order line $E_0=(m_b^2-
m_h^2)/2m_b^2$:
$\aver{E_{\gamma}-E_0}\simeq {\cal O}(\bar \Lambda ^2/m_b)$,
$\aver{(E_{\gamma}-E_0)^2}\simeq {\cal O}(\bar \Lambda ^2)$, i.e.
the
{\em center} of the photon spectrum is shifted from the original line
upward by a small amount of order $\bar \Lambda ^2/m_b$ with
the spectrum acquiring a {\em spreadth} of order $\bar \Lambda$.
More specifically, in the SV expansion one builds up the energy
spectrum step by step: at
${\cal O}(v^0)$ one has only the elastic line of the free quark picture;
at ${\cal O}(v)$ the elastic line gets shifted upward and at ${\cal
O}(v^2)$
the height of the elastic peak is reduced with the missing
contribution being
re-incarnated as {\em inelastic} contributions due to higher
resonances;
perturbative gluon emission finally generates the radiative tail
representing the continuum contributions from high-mass hadronic
final states. The emerging picture provides us with strong
circumstantial evidence that the $B\ra l \nu X_c$ transition can be
treated in the SV limit: for it is observed that the (quasi-)elastic
channels $B\ra l \nu D/D^*$ make up about two thirds of the total
semileptonic width.

{\em ad (ii):} For the asymptotic mass difference
$\bar \Lambda \equiv (M_B-m_b)_{m_b\ra \infty}$ one derives
$$\bar \Lambda (\mu )=\frac{2}{v^2}\int ^{E_0}_{E_0-\mu } dE_{\ga}
\frac{1}{\Gamma} \frac{d\Gamma }{dE_{\ga}}(E_0 - E_{\ga})
\eqno(16)$$
where $\mu$ denotes the IR scale separating short and long distance
dynamics as introduced through the Wilson OPE. Similarly one finds
$$\aver{(\vec p_b)^2}_B=
\frac{3}{v^2}\int ^{E_0}_{E_0-\mu } dE_{\ga}
\frac{1}{\Gamma} \frac{d\Gamma }{dE_{\ga}}(E_0 - E_{\ga})^2
\eqno(17)$$
A few comments might help to elucidate the meaning of eq.(16):
Gluon emission obviously contributes to the $\gamma$ spectrum in
its entire
domain $0\leq E_{\ga}\leq E_0$. Yet for $E_{\ga}$ very close to
$E_0$
the emitted gluon is soft; it then is part of the nonperturbative
medium where the $b$ quark decays and thus has to be incorporated
into the
matrix element. This again illustrates the need for introducing the IR
cut-off $\mu$ chosen to be not too small, as implied by eqs.(16,17).
At
the same time it is quite conceivable that the numerical dependance
on the
concrete value of $\mu$ is mild \cite{POLEMASS}.

{\em ad (iii):} Consider the excitations in $B\ra l\nu X$ produced by
the vector current. For $m_l=0$ there is no elastic channel. Positivity
constraints applied at the point of zero recoil kinematics
then allows to derive -- in a field-theoretical way -- an inequality
previously obtained through a quantum-mechanical line of
reasoning:
$$\aver {(\vec p_b)^2}_B \geq \aver {\mu _G^2}_B\eqno(18)$$

{\em ad (iv):} HQET provides us with an intriguing way to extract the
KM
parameter $|V(cb)|$ from exclusive semileptonic $B$ decays. The
prescription involves two steps \cite{SV2,WISGUR}:
($\al$) One measures $B\ra l \nu D^*$ decays and extrapolates
to zero-recoil, thus determining $|F_{B\ra D^*}(0)V(cb)|$; an
average over the most
recent analyses of CLEO, ALEPH and ARGUS data yield \cite{CLEO2} :
$$ |F_{B\ra D^*}(0)V(cb)| = 0.0367 \pm 0.0025
\eqno(19)$$
($\beta$) For $m_b,\, m_c\ra \infty$ one has $|F_{B\ra D^*}(0)|=1$;
yet at
finite quark masses one has deviations from this symmetry limit, i.e.
$ |F_{B\ra D^*}|(0)=1+{\cal O}(\al _S/\pi ) + {\cal O}(1/m_c^2)
+ {\cal O}(1/(m_bm_c))+{\cal O}(1/m_b^2)$.
It is important to notice that the scale of the
nonperturbative corrections is set by the inverse of the
{\em smaller} mass, i.e. the charm
mass. Thus one expects corrections
$\sim (\mu _{had}/m_c)^2\simeq 0.1$ rather than the previously
claimed
value $\simeq 0.02$ \cite{NEUBERT2} which would correspond to
$(\mu _{had}/m_b)^2$.
Indeed a SV sum rule yields \cite{OPTICAL}
$$1-F^2_{B\ra D^*}(0)=\frac{1}{3}\frac{\aver{\mu _G^2}_B}{m_c^2}+
\frac{\aver{(\vec p_b)^2}_B- \aver{\mu _G^2}_B}{4}
\left(
\frac{1}{m_c^2}+ \frac{1}{m_b^2} + \frac{2}{3m_cm_b}\right)
+ \sum F_{B\ra excitat.}^2 \eqno(20)$$
Due to eq.(18) and the positivity of the individual contributions to
the inclusive rate we see that there is thus an upper bound on the
formfactor
for the quasi-elastic exclusive channel $B\ra l\nu D^*$.
\footnote{Doubt has recently been expressed concerning the validity
of this sum rule; yet that criticism is based on gross misconceptions
about the derivation.} Including
perturbative corrections omitted in eq.(20) we find
$|F_{B\ra D^*}(0)|< 0.94$ as a model-independant upper bound; using
a value for $\aver{(\vec p_b)^2}_B$ as deduced from QCD sum rules
and making a
reasonable allowance for the inelastic channels we arrive at
$|F_{B\ra D^*}(0)|\simeq 0.90 \pm 0.03$, where our guestimate of the
uncertainty reflects the fact that terms $\sim {\cal O}(1/m_c^3)$
have been ignored.
Eq.(19) then gets translated into \cite{SUV}
$$ |V(cb)|_{excl} \simeq 0.0408 \pm 0.003|_{experim}
\pm 0.002|_{theor} \eqno(21)$$
with the first error being experimental and the second one
representing the theoretical uncertainty.
Clearly one has to be tremendously pleased (and relieved) that the
inclusive
and the exclusive analysis, eq.(15) and eq.(21),  yield perfectly
consistent values for $|V(cb)|$.

\section{Extracting $|V(ub)/V(cb)|$}

In Sect.3 I had already stated that at present there
exist practical limitations for directly computing the lepton
energy spectrum so close to the endpoint as to allow an
extraction of $|V(ub)/V(cb)|$. Yet even without a future
breakthrough in computational prowess there are various promising
avenues to -- or at least towards  -- that goal:

\noindent (i) To prepare the ground one extracts the parameters
$\bar \Lambda$ and $\aver{(\vec p_b)^2}_B$ from the first and
second moments of the lepton spectrum in $B\ra l \nu X_c$
transitions \cite{GROZIN}, in analogy to the discussion of eqs.(16,17).

\noindent (ii) With the $b$ quark distribution function
determined through a measurement of the $\gamma$ spectrum in
$B\ra \gamma X$ transitions, one can express the lepton spectrum in
$B\ra l\nu X_u$ as a function of $|V(ub)/V(cb)|$
\cite{MOTION,NEUBERT}.

\noindent (iii) One can fit
the refined $AC^2M^2$ model (as sketched in Sect.3) to the lepton
spectrum beyond the kinematical boundary for $B\ra l \nu X_c$.  In
ref.\cite{ACMREFINED}
the distribution function has been given in terms of the single
parameter
$\aver{(\vec p_b)^2}_B$. The fit would then also return a value for
the
expectation value for the kinetic energy which could then be
compared
with other determinations of this quantity.

\noindent (iv) However one has to face the following complication:
beyond order $1/m_b^2$ there arise non-spectator contributions to
$B\ra l \nu X_u$ transitions; at order $1/m_b^3$ they affect charged,
but
not neutral $B$ decays \cite{WA}. While their contribution to the
total width is small,
it is concentrated in the endpoint region. Ignoring it could then very
seriously affect the value extracted for $|V(ub)/V(cb)|$. To isolate
this effect
one has to measure the endpoint spectrum in charged and neutral
$B$ decays separately. This is a steep price, yet presumably
unavoidable if one
wants to reduce the theoretical error in the value extracted for
$|V(ub)/V(cb)|$ below the 10-20 \% level.

\section{Summary and Outlook}

Our theoretical understanding of inclusive heavy flavour decays has
been advanced tremendously over the last few years,
and we have every reason to expect progress to continue for some
time.
Let me repeat just a few salient points in support of this thesis:

\noindent $\bullet$ Important and illuminating insights into the
workings
of QCD have been gained. Discussing the impact of renormalons onto
heavy flavour decays nicely illustrates the level of technical
sophistication that has become state-of-the-art in this field.

\noindent $\bullet$ New opportunities for cooperation between the
various
Post-Voodoo theoretical technologies have been identified and are
being
exploited now. It represents merely a speed bump in the road
towards progress that such opportunities for cooperation are often
first seen as areas of confusion or even outright conflict.

\noindent $\bullet$ Significant practical gains are being made as
well:

\noindent -- Certain questions that only two or three years ago were
beyond the scope of a theoretical discussion and could be addressed
only phenomenologically can now be treated in a meaningful way,
like whether the semileptonic $B$ branching ratio is 11\% or 13\%;
whether the lifetimes of charged and neutral $B$ mesons differ by
less than 10\% or by more; what the theoretical uncertainty in the
extraction of$|V(cb)|$ is, etc.

\noindent -- $|V(cb)|$  has been extracted in two systematically
quite different ways, namely in {\em inclusive} and in {\em
exclusive} semileptonic $B$ decays, yielding consistent values with a
theoretical uncertainty that in both cases does not exceed the
experimental error:
$$|V(cb)|_{incl} = 0.0410 \pm 0.002|_{experim} \pm 0.002|_{theor}$$
$$|V(cb)|_{excl} = 0.0408 \pm 0.003|_{experim} \pm 0.002|_{theor}$$
This is a major theoretical success that should be savoured
appropriately \footnote{The fact that the two central values agree so
well should {\em not} be overinterpreted; for the stated
uncertainties have at present to be considered as realistic and not
overly conservative.}. Naturally we want to do even better in the
future. My own feeling is that the
{\em inclusive} method offers a higher potential for reducing the
theoretical uncertainties, mainly because it requires `only' an
extrapolation of existing computational techniques coupled with a
measurement of $m_b=M_B-\bar \Lambda$ and
$\aver{(\vec p_b)^2}_B$ as described above. For treating the {\em
exclusive} transition with better accuracy a computational
breakthrough has to be achieved to control the (non-local)
contributions from the higher excitations; this is made apparent by
the analysis of ref.\cite{MANNEL}. Contrary to a recent claim
\cite{NEUBERTPERT} the perturbative corrections are under control
in both the inclusive and exclusive transition \cite{SU}.

\noindent -- The observation that in many instances perturbative
rather than nonperturbative corrections represent the main
theoretical uncertainty illuminates in a nutshell the progress we
have achieved. The size of perturbative corrections can be
determined definitively only through a two-loop calculation; that has
not been done yet. The BLM approach suggests \cite{BLM} that the
$\as ^2$ term is uncomfortably large for inclusive semileptonic $B$
decays. Yet it is pointed out in ref.\cite{SU} that this approach is
quite misleading in heavy flavour decays.

\noindent $\bullet$ New directions for further research have been
identified whose relevance and promise had not been clear a priori.
The SV sum rules can be cited as one recent example of theoretical as
well
as experimental interest. But there are many others:

\noindent -- Analysing $B\ra l\nu D^*$ for extracting $|V(cb)|$;

\noindent -- Measuring the photon spectrum in radiative $B$ decays
to deduce the motion of the $b$ quark inside $B$ mesons;

\noindent -- Observing $B\ra \gamma D^*,\, l^+l^-D^{(*)}$,
$D\ra \ga K^*$ to isolate non-Penguin contributions to radiative $B$
decays;

\noindent -- Studying partially integrated lepton spectra to
determine the mass of the heavy quark and its kinetic energy;

\noindent -- Comparing the endpoint spectra in the semileptonic
decays of charged and neutral $B$ mesons to obtain a more reliable
extraction of
$|V(ub)/V(cb)|$.

All of this does not mean, of course, that we have all the answers or
will obtain them in a straightforward way; we will encounter
unpleasant surprises. Yet it does mean that the theoretical
treatment of heavy flavour decays represents no longer an
embarrassment!

{\bf Acknowledgements}: It is a special pleasure to thank the
organizers of
the meeting, in particular A. Sanda and M. Suzuki, for creating such a
stimulating and efficiently run meeting with a nicely balanced
agenda and further enrichment by cultural entertainment. I am
deeply grateful to my collaborators for
patiently and generously sharing their wisdom with me. This work
was supported in part by the National Science Foundation under
grant number PHY 92-13313.
%
% REFERENCES
%

\end{document}